\documentclass[%
reprint,
superscriptaddress,
showpacs,
amsmath,amssymb,
aps,
prb,
]{revtex4-1}
\usepackage{mathtools}
\usepackage{graphicx,epstopdf}
\usepackage{placeins}
\usepackage[hidelinks]{hyperref}
\usepackage{tikz}
\usepackage{tikz-3dplot}
\usetikzlibrary{calc,math}
\usetikzlibrary{intersections}
\usetikzlibrary{positioning}
\usetikzlibrary{arrows,decorations.markings}
\usepackage{pgfplots}
\pgfplotsset{compat=newest}
\usepgflibrary{shapes.geometric}
\usepackage{mathrsfs}
\usepackage{bm}
\usepackage{esint}
\usepackage{siunitx}
\usepackage{physics}
\usepackage{circledsteps}
\usepackage{yhmath}
\usetikzlibrary{arrows.meta}
\usepackage{relsize}
\definecolor{blue}{RGB}{0,0,200}

\begin{document}
%	\ioptwocol[{
\title{Spinor formulation of the Landau–Lifshitz–Gilbert equation with geometric algebra}
\author{Kristjan Ottar Klausen}
\email{kristjank@hi.is}
\author{Snorri Ingvarsson}

\address{Science Institute, University of Iceland, Dunhaga 3, IS-107 Reykjavik, Iceland.}
\vspace{10 mm}

\begin{abstract}
The Landau-Lifshitz-Gilbert equation for magnetization dynamics is recast into spinor form using the real-valued Clifford algebra (geometric algebra) of three-space. We show how the undamped case can be explicitly solved to obtain component-wise solutions, with clear geometrical meaning.  Generalizations of the approach to include damping are formulated. The implications of the axial property of the magnetization vector are briefly discussed. 

\end{abstract}
\vspace{-20 mm}
\maketitle
\section{INTRODUCTION}
Recent advancements in fabrication and theoretical understanding of spin-dependent heterostructures near nanoscale have opened up a plethora of new technological abilities, applicable in the data and energy sectors \cite{Review_Hillebrands,Review_Sierra}. Short time scales of spin flipping and low energy cost of spin transport allow for the fabrication of faster next-generation technological components, with significantly lowered power consumption compared to conventional ones \cite{Review_Rajput}.
Strong spin-orbit coupling in various condensed matter systems plays a key role and has allowed for the realization of multiple types of quasi-particles, which are stable fermion-like excitations, such as skyrmions \cite{Muhlbauer_2009} and Majorana zero modes \cite{Review_Agauado}. Both have topological properties and hold promise in next-generation information processors, skyrmions as low-power memory and logic devices \cite{Marrows_2021} and Majorana zero modes as qubits for quantum computation \cite{Sarma2015}.

The basic dynamics of both spin and magnetization can be described by the Landau-Lifshitz-Gilbert (LLG) equation \cite{Landau_1935,Gilbert_04}, a rich non-linear equation capable of expressing multiple dynamical structures emergent in systems of angular momenta, such as spin waves and solitons \cite{Lakshmanan_2011}. Additionally, the equation has a mechanical analogue in the kinematical description of a spinning top in a gravitational field \cite{Wegrowe_2012}. The standard form of the equation is commonly formulated with vector calculus, leading to multiple terms of cross products, that can be cumbersome to work with, especially when additional damping and driving terms are included \cite{Meo_2022}. 

Geometric algebra, a real-valued formulation of the more studied Clifford algebras\cite{Floerchinger_21,Garling_2011,Lounesto_2001,Porteous_1995}, has shown to be an accessible generalization of vector and matrix algebra, useful in describing rotations in multiple dimensions \cite{Doran_Lasenby_2003,CA2GC,MacDonald2010_linear}. By incorporating an exterior wedge product, generalizing the cross product, axial vectors are replaced by bivectors, which can form closed algebras isomorphic to the spin symmetry groups \cite{LieGroupsSpinGroups,Lawson2016spin}.

In this work, we apply geometric algebra to rewrite and solve the undamped Landau-Lifshitz-Gilbert equation, and further formulate general expressions including damping. To make the work self-contained, we start by briefly introducing geometric algebra of three dimensional space, whilst also noting some important subtleties regarding its relations to other formalisms. The standard form of the undamped LLG equation is then rewritten and solved with geometric algebra, and approaches to the damped case suggested. The results pave the way for implementation of geometric algebra in condensed matter physics.

\section{Geometric algebra of three-space}
Given a basis of three orthonormal unit vectors $\{\mathbf{e}_1,\mathbf{e}_2,\mathbf{e}_3\}$, the geometric algebra of three-space, $\mathbb{G}_3$, can be formulated by considering an antisymmetric outer product, in addition to a symmetric inner product, combined in a single operator of vector multiplication, known as the geometric product,
\begin{equation}
	\mathbf{ab= a\cdot b + a\wedge b} \ .
	\label{geoprod}
\end{equation}
The symmetric inner (scalar) product has the property
\begin{equation}
	\mathbf{a\cdot b} = \frac{1}{2} (\mathbf{ab} + \mathbf{ba}),
\end{equation}
whilst the anti-symmetric outer (wedge) product satisfies
\begin{equation}
	\mathbf{a} \wedge \mathbf{b}= \frac{1}{2} (\mathbf{ab} - \mathbf{ba}).
\end{equation}
The geometric product of orthogonal vectors results solely in an anti-symmetric component, termed a bivector or a grade two vector, for example,
\begin{equation}
	\mathbf{e}_1 \mathbf{e}_2 = \mathbf{e}_1 \cdot \mathbf{e}_2 + \mathbf{e}_1 \wedge \mathbf{e}_2 = 0 + \mathbf{e}_1 \wedge \mathbf{e}_2 \equiv \mathbf{e}_{12},
\end{equation}
defining $\mathbf{e}_{12}$.
To distinguish vectors from bivectors, we denote the former from here on out with $\vec{v}$ and the latter with $\wideparen{v}$. Working within $\mathbb{G}_3$ with the basis $\vec{e}_i$ for $i=1,2,3$, the algebra has the following elements: scalars, three vectors $\vec{e}_i$, three bivectors $\wideparen{e}_{ij}=\vec{e_i} \wedge \vec{e}_j$, and the trivector pseudoscalar $I=\vec{e}_1 \wedge \vec{e}_2 \wedge \vec{e}_3$. The basis vectors satisfy orthogonality conditions $|\vec{e}_i \cdot \vec{e}_j| =\delta_{ij}$ and $|\vec{e}_i \wedge \vec{e}_j |=1-\delta_{ij}$.

The product of two bivectors has three components in general, a scalar, bivector and quadvector,
\begin{equation}
	\wideparen{A}\,\wideparen{B}= \wideparen{A} \cdot \wideparen{B} + \wideparen{A} \times \wideparen{B} + \wideparen{A} \wedge \wideparen{B},
\end{equation}
where 
\begin{equation}
	\wideparen{A} \times \wideparen{B} = \frac{1}{2} [\wideparen{A}\wideparen{B} -\wideparen{B}\wideparen{A}]
	\label{com}
\end{equation}
denotes the grade-preserving antisymmetric commutator product. When confined to a three component basis, the quadvector term is zero and the geometric product of bivectors has two components, just like the geometric product of vectors in Eq.\ \eqref{geoprod}. For this reason, the exterior algebra of bivectors in three dimensions is structurally similar to the standard vector algebra in three dimensions using the cross product, as axial vectors are dual to bivectors, although their physical interpretations differ. Additionally, the geometric algebra of three-space is isomorphic to the algebra of Pauli matrices\cite{Hestenes_STA}.

\begin{figure}[t!]
	\centering
	\includegraphics[width=0.45\textwidth]{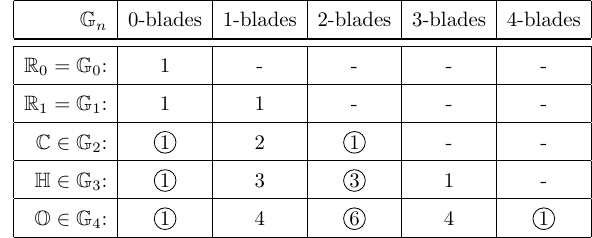}
	\caption{Binomial expansion of geometric algebra up to $\mathbb{G}_4$. The hypercomplex division algebras of complex numbers ($\mathbb{C}$), quaternions ($\mathbb{H}$) and octonions ($\mathbb{Q}$) are shown (circled) as the even component subsets of the respective geometric algebra.}
	\label{binom}
\end{figure}

Since the product of orthogonal bivectors in three-space results in a bivector, they form a closed subalgebra, isomorphic to the quaternions. Moreover, the even components of geometric algebra in each dimension form a closed subspace, Fig.\ \ref{binom}. 
In two-dimensional geometric algebra, the even subalgebra consists of scalar and bivector-valued pseudoscalar components, $x+iy$ for $x,y \in \mathbb{R}$ where $i=\vec{e}_1 \wedge \vec{e}_2$ can be considered as the unit imaginary. To distinguish this bivector-valued psuedoscalar of the plane from the trivector-valued pseudoscalar of space, we denote the latter with $I=\vec{e}_1 \wedge \vec{e}_2 \wedge \vec{e}_3$.

It has been shown that every Lie algebra is isomorphic to a bivector algebra, and every Lie group can be represented as a spin group \cite{LieGroupsSpinGroups}. In this way, spinors can be understood as elements of an even subalgebra of the corresponding geometric algebra \cite{Francis_2005}.

%In general, for an $r$-vector and $s$-vector, where $m=\frac{1}{2} r+s - |r-s|$, the geometric product is given by\cite{CA2GC}
%\begin{align}
%	&A_r B_s = \sum_{k=0}^{m} \langle A_r B_s \rangle_{|r-s| + 2k} \\
%	&= \langle A_r B_s \rangle_{|r-s|} + \langle A_r B_s \rangle_{|r-s|+2} + ... + \langle A_r B_s. \rangle_{|r-s|+2k}
%\end{align}

\section{The Landau-Lifshitz equation}
\label{Sec:LL}

\begin{figure}[t!]
	\centering
\includegraphics[scale=1]{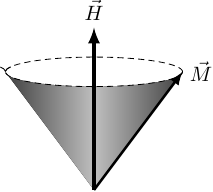}
	\caption{Precession of the magnetization vector, $\vec{M}$, in an external magnetic field, $\vec{H}$.}
	\label{LLfig}
\end{figure}

Precession without damping of the magnetization vector $\vec{M}$ in a magnetic material, subjected to an external magnetic field $\vec{H}$, Fig.\ \ref{LLfig}, can be described with vector calculus by the Landau-Lifshitz (LL) equation
\begin{equation}
	\frac{\partial \vec{M}}{\partial t}= -\gamma \vec{M} \times \vec{H},
	\label{LL}
\end{equation}
where $\gamma$ is the gyromagnetic ratio, $$\gamma= \mu_0 \gamma_e=\frac{g e  \mu_0}{2m},$$ in which $g$ denotes the g-factor, $\mu_0$ is the vacuum permeability, $m$ and $e$ are the electron mass and charge respectively. 
Validity of the equation is well established from nearly a century of experiments in magnetics \cite{Landau_1935}. Applying methodology from Ref.\ \onlinecite{NFCM}, we proceed to rewrite and solve the undamped LL-equation with geometric algebra.

\begin{figure}[b!]
	\includegraphics[width=0.35\textwidth]{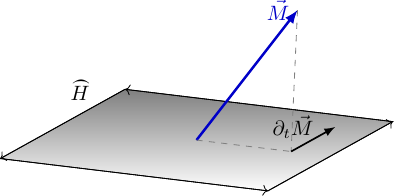}
	\caption{Conceptual diagram of the inner product between a vector $\vec{M}$ and bivector $\wideparen{H}$ resulting in the vector $\partial_t\vec{M}$ (arbitrary size scaling).}
	\label{int_bi_vec}
\end{figure}

Using the duality between the cross product and wedge product in three dimensions, expressible with the pseudoscalar, $\vec{a} \times \vec{b}= I (\vec{a} \wedge \vec{b})$, along with the property $I (\vec{a} \wedge \vec{b})= \vec{b} \cdot (I\vec{a})$, the equation can be rewritten as

\begin{equation}
	\frac{\partial \vec{M}}{\partial t}= -\gamma I (\vec{M} \wedge \vec{H}) =\gamma \vec{M}\cdot I\vec{H} = \gamma \vec{M} \cdot \wideparen{H},
	\label{LLdot}
\end{equation}
resulting in an inner product between the magnetization vector, and the magnetic field bivector $\wideparen{H}$. To get a feel for how the inner product results in the vector $\partial_t \vec{M}$, consider the component of $\vec{M}$ projected onto the plane of $\wideparen{H}$ to cancel (dot out) the corresponding vector component of $\wideparen{H}$, leaving the orthogonal component, Fig.\ \ref{int_bi_vec}. 
Note that the inner product between a vector and bivector is antisymmetric, 
\begin{equation}
	\vec{M} \cdot \wideparen{H}=-\wideparen{H} \cdot \vec{M},
	\label{MHasym}
\end{equation}
whilst the outer product is symmetric in this case (unlike the case for two vectors),
\begin{equation}
 \vec{M}\wedge \wideparen{H}=\wideparen{H}\wedge\vec{M}.
 \label{MHsym}
\end{equation}
With these properties, Eq.\ \eqref{LLdot} can be rewritten further, using the shorthand notation $\partial_t \vec{M}$ for the partial time derivative,
\begin{align}
		&\partial_t \vec{M}- \gamma \vec{M} \cdot \wideparen{H} = 0,\\[2mm]
	&	\partial_t \vec{M}- \frac{\gamma}{2}  \vec{M} \cdot \wideparen{H} + \frac{\gamma}{2}  \wideparen{H} \cdot \vec{M} =0,\\[2mm]
	&	\partial_t \vec{M}- \frac{\gamma}{2}  \vec{M} \wideparen{H} + \frac{\gamma}{2}  \wideparen{H} \vec{M} =0.
\end{align}
This expression can now be written in the form
\begin{equation}
	\frac{d}{dt}\left(R \vec{M} R^\dagger \right)=0,
	\label{ddtLL}
\end{equation}
where $R$ is a spinor operator\cite{Lounesto_2001} that satisfies
\begin{equation}
	\frac{dR}{dt}= \frac{\gamma}{2} R \wideparen{H}.
	\label{spinoreq}
\end{equation}
Conversely, 
\begin{equation}
	\frac{dR^\dagger}{dt}= -\frac{\gamma}{2} \wideparen{H} R^\dagger,
\end{equation}
since
\begin{align*}
	\frac{d}{dt}\left(R \vec{M} R^\dagger\right) &= \frac{dR}{dt} \vec{M}R^\dagger + R \partial_t \vec{M} R^\dagger + R \vec{M} \frac{dR^\dagger}{dt}\\
	&= R \frac{\gamma}{2} \wideparen{H} \vec{M}R^\dagger +  R \partial_t \vec{M} R^\dagger -R \frac{\gamma}{2} \vec{M} \wideparen{H}R^\dagger\\
	&= R\left(\frac{\gamma}{2}  \wideparen{H} \vec{M} + 	\partial_t \vec{M}- \frac{\gamma}{2}  \vec{M} \wideparen{H}\right)R^\dagger =0.
\end{align*}
The Landau-Lifshitz equation, Eq.\ \eqref{LL}, can now be solved by solving the simpler spinor equation, Eq.\ \eqref{spinoreq}. For a constant magnetic field, the solution to \eqref{spinoreq} is
\begin{equation}
	R=\exp\left(\frac{\gamma}{2} \wideparen{H} t\right),
	\label{Rspin}
\end{equation}
and similarly,
\begin{equation}
	R^\dagger=\exp\left(-\frac{\gamma}{2} \wideparen{H} t\right).
\end{equation}
Therefore,
\begin{equation}
	RR^\dagger =1.
	\label{RR}
\end{equation}
For this reason $R$ can be considered as a unit spinor, but is generally referred to as a rotor in geometric algebra \cite{Doran_Lasenby_2003}. In general, for a unit bivector $\wideparen{B}$ we have the corresponding rotor 
\begin{equation}
	S=\cos(\theta) \pm \wideparen{B} \sin(\theta) = \exp(\pm \wideparen{B} \theta) = \exp(\pm I\vec{B}\theta).
	\label{RotorGen}
\end{equation}

\begin{figure}[h!]
	\centering
	\includegraphics[scale=1]{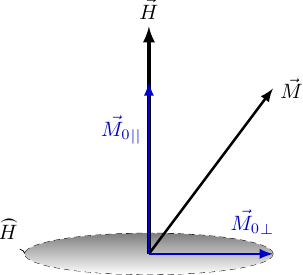}
	\caption{Component solutions of the precession of the magnetization vector $\vec{M}$ in an external magnetic field $\vec{H}=I \wideparen{H}$. The magnetization vector component $\vec{M}_{0\perp}$, perpendicular to the magnetic field vector, $\vec{H}$, is rotated in the plane of the bivector $\wideparen{H}$, whilst the parallel component, $\vec{M}_{0\parallel}$, is constant.}
	\label{LL2}
\end{figure}

Equation \eqref{ddtLL} states that the rotation of $\vec{M}$ in the plane of $\wideparen{H}$ is constant, integrating and using the initial conditions $R(0)=1=R^\dagger(0)$, the magnetization is given by
\begin{equation}
	R \vec{M} R^\dagger = \vec{M}_0,
\end{equation}
where $\vec{M}_0$ denotes the initial magnetization vector, $\vec{M}_0=\vec{M}(0)$. 
Using Eq.\ \eqref{RR}, the solution for $\vec{M}$ can be obtained since $R$ is known from Eq. \eqref{Rspin},
\begin{align}
	\vec{M}(t) &= R^\dagger \vec{M}_0 R\\
	&= e^{-\frac{\gamma}{2} \wideparen{H} t} \vec{M}_0 \, e^{\frac{\gamma}{2} \wideparen{H} t}.
\end{align}
A more revealing form can be found by splitting $\vec{M_0}$ into components parallel and orthogonal to the magnetic field vector $\vec{H}$ respectively, using the properties from Eqs.\ \eqref{MHasym}, \eqref{MHsym}, and \eqref{RotorGen},
\begin{align}
	\vec{M}&=R^\dagger\left(\vec{M_0}_{||} + \vec{M_0}_{\perp}\right)R\\
	&= R^\dagger \vec{M_0}_{||} R + R^\dagger  \vec{M_0}_{\perp} R\\
	&= \vec{M_0}_{||} R^\dagger R +  \vec{M_0}_{\perp} RR\\
	&= \vec{M_0}_{||} + \vec{M_0}_{\perp} R^2\\
	&= \vec{M_0}_{||} + \vec{M_0}_{\perp}  e^{\gamma \wideparen{H} t} .
\end{align}
This shows explicitly that the component of $\vec{M}_0$ parallel to $\vec{H}$ is constant while the orthogonal component rotates in the plane of the bivector $\wideparen{H}$, or equivalently about the axis $\vec{H}=I \wideparen{H}$. To be consistent with the notation of the more classical vector calculus solutions, we have used the magnetic field vector for reference in defining parallel and perpendicular components of $\vec{M}_0$, rather that the dual bivector, Fig.\ \ref{LL2}.

\section{The Landau-Lifshitz-Gilbert equation}
In ferromagnetic materials, the precession of the magnetization vector is damped, in the sense that $\vec{M}$ spirals toward a minimum energy configuration parallel to the magnetic field vector $\vec{H}$. Landau and Lifshitz\cite{LL_51} introduced a phenomenological damping parameter,  $\lambda$, to describe this.
Including damping, the LL equation reads 
\begin{equation}
	\frac{\partial \vec{M}}{\partial t}= -\gamma \vec{M} \times \vec{H} - \lambda \vec{M} \times (\vec{M} \times \vec{H}),
	\label{LLl}
\end{equation}
where $\lambda$ has the dimension of $TL^2/Q^2$ in terms of time (T), length (L) and charge (Q). The damping parameter is commonly written with $\lambda=\lambda'/M_s^2$ where $M_s$ is the saturation magnetization, in which case $\lambda'$ has dimension of frequency, or $1/T$.
The damping term can be rewritten by substitution with Eq.\ \eqref{LL} to a form expounded by Gilbert \cite{Gilbert_04}, known as the Landau-Lifshitz-Gilbert equation (LLG),
\begin{equation}
	\frac{\partial \vec{M}}{\partial t}= -\gamma \vec{M} \times \vec{H} + \alpha \vec{M} \times \frac{\partial \vec{M}}{\partial t},
	\label{LLG}
\end{equation}
with $\alpha= \alpha'/M_s$ where $\alpha'=\lambda' /M_s \gamma = \lambda M_s/\gamma$ is a dimensionless parameter. This form of the damped equation can be motivated in a Lagrangian approach by the addition of a Rayleigh dissipation function, a quadratic function of dynamical variables commonly applied to include velocity-dependent frictional forces \cite{Gilbert_04}.

Applying comparable methodology as in Sect.\ \ref{Sec:LL}, the LLG equation \eqref{LLG} can be rewritten in a spinor from, denoting the damping vector with $\vec{g}(t)=\alpha \vec{M} \times \partial_t \vec{M}$ for readability,
\begin{equation}
	\frac{d}{dt} (R\vec{M}R^\dagger)= R \vec{g} R^\dagger,
	\label{RgRd}
\end{equation}
where $R=\exp(\frac{\gamma}{2}\wideparen{H}t)$. The spinor form, Eq.\ \eqref{RgRd}, can be interpreted to state that the plane of rotation for the magnetization vector $\vec{M}$ is changed by the damping vector, which itself is rotated by the external magnetic field. The general solution is then 
\begin{equation}
	\vec{M}(t)= R^\dagger \left( \vec{M}_0 + \int_{0}^{t} R \vec{g}(t') R^\dagger dt'\right)R.
\end{equation}
This functional equation can be analyzed further by looking at parallel and orthogonal components of the solution, and simplified by making case-specific approximations to the relations between $\vec{M}$, $\vec{H}$ and $\vec{g}$.

An alternative approach is to rewrite Eq. \eqref{LLG} as
\begin{equation}
	\frac{\partial \vec{M}}{\partial t} = -\gamma \vec{M} \times \vec{H}_{e},
\end{equation}
using the effective magnetic field $\vec{H}_e= \vec{H}_0 - \eta \vec{H}_d$ where $\vec{H}_d (t)= \partial_t \vec{M}(t)$ and $\eta = \alpha/\gamma$. Proceeding analogously as in Sect.\ \ref{Sec:LL}, one can then consider the equivalent spinor equation
\begin{equation}
	\frac{d}{dt} (U\vec{M}U^\dagger)=0,
\end{equation}
where the spinor $U$ satisfies
\begin{equation}
	\frac{dU}{dt} = \frac{\gamma}{2} \wideparen{H}_e U = \frac{\gamma}{2} I (\vec{H}_0 - \eta \vec{H}_d) U.
\end{equation}
The composite spinor $U=\exp(\frac{\gamma}{2} \wideparen{H}_0 t - \frac{\alpha}{2} I\vec{M}(t))$ is a solution and the magnetization is given by
\begin{equation}
	\vec{M}(t) = U^\dagger \vec{M}_0 U.
\end{equation}
The recursive non-linearity is still implicit in the two suggested approaches, as the  solution for the magnetization is a function of itself. 
Another approach is to factor the composite spinor $U$ such that $U=RS$. However, care must be taken in the analysis as the bivectors $\wideparen{H}$ and $I\vec{M}$ are generally not orthogonal, in which case the anti-commutative property of the bivectors comes into play \cite{Doran_Lasenby_2003}.\\

\section{DISCUSSION}
The Landau-Lifshitz-Gilbert equation is routinely applied to describe uniform precession observed in ferromagnetic resonance experiments\cite{Heinrich03}. It can be used to fit the magnetic susceptibility\cite{Ingvarsson02} and extract parameters such as the magnetic saturation, Gilbert damping coefficient as well as probing the anisotropic energy landscape\cite{Ingvarsson04}. The value of the current approach is simplification of calculations and ease of geometric interpretation. Furthermore, it provides a basis for the analysis of additional torque terms and spin-transport effects\cite{Hellman2017}, such as spin transfer torques\cite{RalphStiles08} and spin pumping terms\cite{Tserkovnyak05}, which can be included in the LLG equation and treated with geometric algebra in a similar manner. This work is currently underway. Here we have shown how torque terms can be rewritten to allow for different methods of solution. Another benefit of geometric algebra is the embedding within the larger algebra\cite{Gurtler75}, with more degrees of freedom, that may prove helpful in describing the various magnetization effects arising in coupled systems, calling for additional terms in the LLG equation \cite{Quarenta24,Verstraten23,Atxitia17,Zhang09}.

In the preceding analysis, the magnetization has been described with a vector. However, just as the magnetic field is an axial vector, more suitably described by it's dual bivector, the same applies to the magnetization. For the sake of familiarity and similarity to traditional approaches, we have chosen to keep the vector form of the magnetization in the current work.
By considering the magnetization as a bivector, the antisymmetric product with the magnetic field bivector takes the form of a commutator, establishing similarity to quantum mechanical treatments of angular momentum. To further apply geometric algebra in the analysis, space-time can be incorporated by an additional temporal basis vector\cite{Hestenes_STA}.\\

\section{CONCLUSIONS}
Magnetization dynamics is governed by the Landau-Lifshitz-Gilbert equation, which employs vector calculus to describe the relations between the magnetic torques. 
By rewriting the equation with geometric algebra, we have shown how component-wise solutions can be readily obtained and visualized in the undamped case. Furthermore, two approaches to a general solution of the damped case have been formulated. Geometric algebra enables the use of a spinor form of the equations of motion, simplifying calculations whilst retaining ease of interpretation. The formalism lies much closer to quantum mechanical treatments of spin dynamics, and can provide a fruitful platform for analyzing spin dynamics in general. For the sake of similarity to standard approaches, the scope of current work is restricted by the sole consideration of the external magnetic field as a bivector whilst keeping the magnetization as a vector.

Although the implementation of geometric algebra is well underway in robotics, computer science, gravitation and quantum mechanics \cite{Hitzer2022}, applications is condensed matter physics are still few. The current work is a step into this domain.

\begin{acknowledgements}
	This work was supported by funding from the Icelandic Research Fund, grants no. 239623 and 228951.
\end{acknowledgements}

\bibliographystyle{unsrt} %byrjar á 1,2...
\bibliography{ref_LLG_GA}
\end{document}